\documentstyle[12pt,amsfonts,amssymb]{article}

\begin{document}
\hskip 13truecm{ETH-TH/95-7}
\smallskip

\hskip 13truecm{February 1995}
\vskip 1truecm
\centerline{\large{Conservation Laws and Formation of Singularities}}
\centerline{\large{in Relativistic Theories of Extended Objects}}
\vskip .5truecm
\centerline{Jens
Hoppe\renewcommand{\thefootnote}{\fnsymbol{footnote}}\footnote[1]{Heisenberg
Fellow\\
On leave of absence from
the Institute for Theoretical Physics, Karlsruhe University,
Germany}\renewcommand{\thefootnote}{\arabic{footnote}}}
\centerline{Institut f\"ur Theoretische Physik}
\centerline{ETH H\"onggerberg}
\centerline{CH-8093 Z\"urich}
\vskip 2truecm
\centerline{\bf{Abstract}}\rm
\vskip 1truecm
The dynamics of an M-dimensional extended object whose M+1 dimensional world
volume in M+2 dimensional space-time has vanishing mean curvature is
formulated in term of geometrical variables (the first and second fundamental
form of the time-dependent surface $\sum_M$), and simple relations involving
the
rate of change of the total area of $\sum_M$, the enclosed volume as well as
the spatial mean -- and intrinsic scalar curvature, integrated over $\sum_M$,
are derived. It is shown that the non-linear equations of motion for
$\sum_M(t)$ can be viewed as consistency conditions of an associated linear
system that gives rise to the existence of non-local conserved quantities
(involving the Christoffel-symbols of the flat M+1 dimensional euclidean
submanifold swept out in ${\Bbb R}^{M+1}$). For M=1 one can show that all
motions are necessarily singular (the curvature of a closed string in the
plane can not be everywhere regular at all times) and for M=2, an explicit
solution in terms of elliptic functions is exhibited, which is neither
rotationally nor axially symmetric.

As a by-product, 3-fold-periodic spacelike maximal hypersurfaces in ${\Bbb
R}^{1,3}$ are found.
\vfill\eject
\noindent\bf{I. Introduction}\rm
\vskip .5truecm
Consider the motion of an M-dimensional extended object $\sum_M(t)$ in
${\Bbb R}^{M+1}$. Any such motion gives rise to a $(M+1)$-dimensional manifold
$\Bbb M$ in $(M+2)$-dimensional space-time ${\Bbb R}^{1,M+1}$,
whose boundaries
(if $\sum_M$ is compact) are $\sum_M$ (initial time $t_i$) and $\sum_M$ (final
time $t_f$). Relativistically invariant dynamics for $\sum_M$ can be
formulated by subjecting ${\Bbb M}$ to a variational principle, like the
extremization of the volume-functional (generalizing [1]). The volume of
${\Bbb M}$ may be given by introducing coordinates
$(\varphi^{\alpha})_{\alpha=0, \cdots, M}$ on ${\Bbb M}$, describing ${\Bbb M}
\subset {\Bbb R}^{1, M+1}$ by the $M+2$ coordinate-functions
$x^{\mu}(\varphi^0, \cdots, \varphi^M)$, calculating the metric
$G_{\alpha\beta}$ induced by the
flat Minkowski-metric $(\eta_{\mu \nu})_{\mu\nu=0, \cdots, M+1}=
\mbox{diag}(1, -1, \cdots, -1)$ and integrating,
\begin{eqnarray}
S & = & \mbox{Vol}({\Bbb M})=\int d\varphi^{M+1}\sqrt G\\
\nonumber G & = & (-)^M\mbox{det}(G_{\alpha\beta}) \qquad , \qquad
G_{\alpha\beta}=\frac{\partial
x^{\mu}}{\partial\varphi^{\alpha}} \, \, \frac{\partial
x^{\nu}}{\partial\varphi^{\beta}}\eta_{\mu\nu} \quad .
\end{eqnarray}

Taking (1) as a starting point (with signature $({\Bbb M})=(1, -1, \cdots,
-1)$)
one may ask: what does extremality of S (considered as a functional of the
$x^{\mu}$) imply for $\sum_M(t)$, the shape of the extended object? Choosing
$\varphi^0=x^0=t$, and the time dependence of the spatial parameters
$\varphi=(\varphi^r)_{r=1, \cdots, M}$ such that the motion of $\sum_M$
(described by $\vec x(t, \varphi)=(x^1, \cdots, x^{M+1})$) is always normal,
i.e.
\begin{equation}
(G_{\alpha\beta})=\left(\begin{array}{cc}
\displaystyle{1-{\dot{\vec x}}^2}
& \displaystyle 0 \cdots 0 \\
0 \\
\vdots &-\partial_r\vec x\cdot\partial_{s}{\vec x}\\
0 \end{array}\right)
\end{equation}
the extremality condition(s)
\begin{eqnarray}
\frac{1}{\sqrt{G}}\partial_\alpha\sqrt{G} G^{\alpha\beta}\partial_{\beta}
x^{\mu}
&= & 0 \\
\nonumber \mu & = & 0, \cdots, M+1
\end{eqnarray}
read:
\begin{equation}
\frac{\partial}{\partial t}\left (\sqrt{g}\sqrt{\frac{1}{1-{\dot{\vec
x}}{^2\,}}}\right )=0
\end{equation}
\begin{equation}\rho\cdot{\ddot{\vec x}}=
\partial_r\frac{1}{\rho}gg^{rs}\partial_x\vec x
\end{equation}
\[
\rho=\rho(\varphi^1, \cdots, \varphi^M):=\sqrt{\frac{g}{1-{\dot{\vec
x}}{^2}}}
\]
where $\cdot = \frac{\partial}{\partial t}$, and $g$ and $g^{rs}$ are the
determinant and inverse, respectively, of the (positive definite) metric
$g_{rs}:=\partial_r\vec x\partial_s \vec x$ on $\sum_M(t)$. The conservation
law (4), ``large area(densitie)s have to slow down, while
small area(densitie)s
speed up'' (anticipating singularities as well as periodicity), encodes almost
the complete dynamical information. To see this, one first notes that on a
fixed compact surface $\sum_M(t=t_i)$ parameters $(\varphi^r)_{r=1,
\cdots, M}$
may be chosen such that the conserved (energy-)density is actually independent
of $\varphi$, i.e.
\begin{equation}
{\dot{\vec x}}\,{^2}+g/\lambda^{2M}=1 \qquad
\end{equation}
\[
\lambda=\mbox{const.}
\]
-- as noted already in [2], (4) then ensures that (6) will hold for all
$t$. Furthermore, as (5) and the orthogonality conditions (cp. (2))
\begin{equation}
{\dot{\vec x}}\partial_r\vec x=0 \qquad , \qquad r=1, ... M
\end{equation}
are invariant under
\begin{equation}
\vec x(t,\lambda)\to \lambda \vec x(\frac{t}{\lambda},\varphi)
\end{equation}
(corresponding to $x^{\mu}\to\lambda x^{\mu}$ in (3)), one could put
$\lambda=1$ in (6), with the understanding, that each motion with
$\lambda\ne1$ can be obtained from a $\lambda=1$ motion via (8). In any case,
one can show that, since (6) and (7), i.e.
\begin{equation}
{\dot{\vec x}}=\pm\sqrt{1-g/\lambda^{2M}}\vec n \qquad ,
\end{equation}
$\vec n$= surface normal, holds (cp[3]), the equations of motion (5) are
automatically satisfied -- apart from points where ${\dot{\vec x}}=0$. As will
be seen in the next section(s) it is convenient to write (9) in the form
\begin{equation}
{\dot{\vec x}}=-\mbox{sin}\theta \; \vec n
\end{equation}
\[
\theta=\theta(t,\varphi^{1}, \cdots \varphi^M) \; \epsilon \; (-\pi/2,+\pi/2)
\]
One should note that choosing the conserved energy density $\rho$ to be
constant on $\sum$ (i.e. independent of $\varphi$) is a matter of convenience,
not necessity; eq. (5) is a consequence of (4) and (7), resp. (10), for any
$\rho$, and in the considerations that will follow one could equally think of
$\sin^2\theta$ as being given by $1-g/\rho^2(\varphi)$, rather than
$1-g/\lambda^{2M}$. Leaving the density $\rho$ unspecified one would keep full
$\mbox{Diff}\sum$ invariance of the equations.

At first sight (9), with the
normal velocity being a (specific) function of the area-density $\sqrt{g}$
(see [4] for a Hamiltonian formulation, and [5] for a general dependence on
$\sqrt{g}$) may look rather simple -- perhaps simpler than the, by now fairly
well understood, mean-curvature flow (defined by letting the normal velocity
be equal to the mean curvature of $\sum_t$ -- thus involving second
derivatives of $\vec x$, rather than first ones); however, certain crucial
techniques available for the mean curvature flow (see e.g. [6]) do
not apply to
(9).
\vskip .5truecm
\noindent{\bf{II. Formulation of the Dynamics of $\sum_M$ in Terms of
Geometrical Variables}}\rm
\vskip .5truecm
The simple first-order form of the dynamics, (10) (resp. (9)), allow one to
easily derive the basic equations,
\begin{equation}
{\dot g}_{rs}=-2\mbox{sin}\theta \; h_{rs}
\end{equation}
\begin{equation}
{\dot h}_{rs}=(\nabla_r\nabla_s-h_{ra}g^{ab}h_{sb}) \sin\theta
\end{equation}
for the components of the metric tensor, and the second fundamental form
\begin{equation}
h_{rs}:=-\partial_{rs}^2\vec x\cdot\vec n \qquad \qquad ;
\end{equation}
$(\nabla_r)$ are the covariant derivatives (with respect to
$\varphi^r$) on $\sum_t$, i.e.
\begin{equation}
\nabla_a\nabla_b f=\partial_{ab}^2f-\gamma_{ab}^c\partial_cf
\end{equation}
\[
\gamma_{ab}^c=\frac{1}{2}g^{cd}(\partial_ag_{bd}+\partial_bg_{ad}-
\partial_dg_{ab})
\]
for any function f: $\sum_t\to{\Bbb R}$. Note that the
gauge-fixing (cp.(6)) has left one with a residual
$\mbox{SDiff}\sum_t$-invariance, i.e. invariance of the equations under
reparametrisations
\begin{equation}
\varphi^r\to{\tilde{\varphi}}^r(\varphi^1, \cdots\varphi^M)\qquad, \qquad
J=\mbox{det}\frac{\partial{\tilde{\varphi}}^r}{\partial\varphi^s}=1
\end{equation}
and that $\theta$ (remember that $\cos^2\theta=g/\lambda^{2M}$) is an
`observable'. Also note that (5), with $\rho=\lambda^M=\mbox{const}$, implies
\begin{eqnarray}
{\ddot{\vec x}}\cdot\vec n=-\cos^2\theta\cdot H\qquad, \qquad
H: & = & g^{rs}h_{rs} \\
\nonumber & = & \mbox{mean curvature}
\end{eqnarray}
(as well as ${\ddot{\vec x}}\partial_r\vec
x=-\frac{1}{2}\partial_r(g/\lambda^{2M})=\sin\theta\cos\theta\partial_r\theta$
-- which is zero at the turning points, ${\dot{\vec x}}(\tilde t, \tilde
\varphi)=0$); taking the time-derivative of $\sin\theta:=\vec n{\dot{\vec x}}$
one obtains
\begin{equation}
{\dot{\theta}}=\cos\theta \;\; H
\end{equation}
(for $\theta\ne0$, this could have been obtained directly from (11),
$-2\sin\theta\cos\theta{\dot \theta}\lambda^{2M}={\dot g}=gg^{rs}{\dot
g}_{rs}=-2g\sin\theta \; H$).
Calculating $-{\ddot x}^{\mu}n_{\mu}, n^{\mu}$ being normal to ${\Bbb M}$
in ${\Bbb R}^{1, M+1}$,
\begin{equation}
n^{\mu}=\left( \begin{array}{c}
\begin{array}{c}
-\mbox{tan}\theta \\ \displaystyle\frac{\vec n}{\cos \theta}
\end{array}\end{array} \right)
\end{equation}
one can check that $\frac{-{\dot \theta}}{\cos\theta}$ is indeed the curvature
of any $\varphi=\mbox{const}$ curve (worldline) in ${\Bbb M}$ (as it should,
according to (17), to add up to zero, with the spatial principal
curvatures). In any case, (11) and (12) imply
\begin{equation}
{\ddot g}_{rs}=\cot\theta{\dot g}_{rs}{\dot
\theta}+\frac{1}{2}{\dot g}_{ra}g^{ab}{\dot g}_{bs}-2\sin\theta
\nabla_r\nabla_s\sin\theta
\end{equation}
(where ${\dot \theta}$, (cp) (17), could be replaced by
$-\frac{1}{2}\cot\theta g^{ab}{\dot g}_{ab}$). Modulo the gauge-fixing, (19)
is equivalent to the original minimal hypersurface equations. Note that only
for $M=1$, where $g_{rs}=\lambda^2\cos^2\theta(t,\varphi)$ yields ${\ddot
\theta}=\theta''$, one has decreased the number of equations. For any $M$,
letting $(T_g)^a_b:=g^{ac}\dot g_{cb}$, they imply the matrix equation
\begin{equation}
{\dot T}_g=-\frac{1}{2}T_g^2-\frac{1}{2}(\cot\theta)^2(TrT_g)T_g-2\sin
\theta\nabla^{\cdot}\nabla_{\cdot}\sin\theta.
\end{equation}
$\;\;$ In order to make all the $\theta$-dependence explicit, one could
insert
\begin{equation}
g_{rs}=\lambda^2(\cos\theta)^{\frac{2}{M}}\bar g_{rs} \; \; \;\bar g=\det\bar
g_{rs}=1
\end{equation}
into (11)/(12), resp. (19) or (20), which then becomes an equation for the
traceless matrix $\bar T=T_g+\frac{2}{M}H\sin\theta \bf{1}:$
on the other hand it is easy to see directly from (11)/(12) that
\begin{equation}
\bar T_b^a:=\bar g^{ac}{\dot{\bar
g}}_{cb}=2\sin\theta(h_b^a-\frac{H}{M}\delta_b^a) \; \; \; \; \; \;
\frac{1}{2}Tr\bar
T^2=\frac{2}{M}\sin^2\theta{\Bbb T}_{r<s}(\kappa_r-\kappa_s)^2
\end{equation}
and that the Weingarten map $T \, : \, T^a_b=g^{ac}h_{cb}=h^a_b$,
whose eigenvalues are the principal curvatures $\kappa_{r\cdot}$
satisfies
\begin{equation}
{\dot T}=(T^2+\nabla^{\cdot}\nabla_{\cdot})\sin\theta \; \; \; .
\end{equation}
Taking the trace of (23), and integrating over $\sum_t$, one finds that
\begin{equation}
\int{\dot
H}\sqrt{g}d^M\varphi=\int(\sum\limits_{r=1}^M\kappa_r^2)\sin\theta
\sqrt{g}d^M\varphi
\; \; .
\end{equation}
As (11) and (12) were derived from (9) (resp. (10)) which describe the
(time-)deformation of embedded hyper-surfaces, solutions $g_{rs}(t):
\sum_t\to{\Bbb R} \, , \, h_{rs}(t): \sum_t\to{\Bbb R}$ will automatically (if
they do so a $t=t_i)$
satisfy the Gauss-equations
\begin{equation}
R_{abcd}=h_{ac}h_{bd}-h_{ad}h_{bc} \; \; ,
\end{equation}
in particular
\begin{equation}
R:=R_{abcd} \, g^{ac}g^{bd}=H^2-TrT^2 \; \; ,
\end{equation}
and the Codazzi equations
\begin{equation}
\nabla_ah_{bc}=\nabla_bh_{ac} \; \; .
\end{equation}
Due to (26), and ${\dot{\sqrt{g}}}=-\sin\theta H\sqrt{g}$ (24) may also be
stated as
\begin{equation}
\frac{d}{dt}\int_{\sum_{t}}H=-\int_{\sum_{t}}R\sin\theta \; \; .
\end{equation}
(27), on the other hand, is useful when considering the evolution of
$Q_m:=Tr\,T^n$ for $n>1$, e.g.
\begin{equation}
\frac{1}{2}{\dot Q}_2=Q_3+T^a_b\nabla^b\nabla_a\sin\theta\qquad .
\end{equation}
Integrating over $\sum$ (using $\nabla^bT_b^a=\nabla^aH$, and
$\triangle\sin\theta=\dot H-Q_2\sin\theta$) yields
\begin{equation}
\int{\dot
R}\sqrt{g}d^M\varphi=2\int(HQ_2-Q_3)\sin\theta\sqrt{g}d^M\varphi\qquad ,
\end{equation}
respectively
\begin{eqnarray}
\frac{d}{dt}\int_{\sum_{t}}R & = & \int_{\sum_{t}}(3HQ_2-2Q_3-H^3)\sin\theta
\\
& = & - \int_{a\ne b\ne c}(\sum\kappa_a\kappa_b\kappa_c)\sin\theta
\end{eqnarray}
(recovering, for $M=2$, a weak form of the Gauss-Bonnet theorem as a
consequence of the dynamical equations). Also note that the rate of change of
the volume enclosed by $\sum_M$, respectively its total area, are given by
\begin{equation}
{\dot \vee}=-\int\sin\theta\cos d^M\varphi \, , \, {\dot
A}=-\int\sin\theta\cos H d^M\varphi
\end{equation}
\vskip .5truecm
\noindent{\bf{III. Zero Curvature Condition and Non-Local Conserved
Quantities.\rm
\vskip .5truecm

The fact that the dynamical equations (5) are automatically satisfied as a
consequence of gauge-fixing conditions, (7), and a conservation law, (4), --
which too can be stated as a condition on the metric of ${\Bbb M}$ -- may also
be used in the following way: Consider hypersurfaces
$\sum_{{t}_{i}},\sum_{{t}_{f}}$ and
motions in between such that for $t_i\leq t \leq t_f$ all points of
the surface have
non-vanishing velocity. The projection of ${\Bbb M}$ onto
${\Bbb R}^{M+1}$ will
then be a euclidean domain ${\Bbb M}_E\subset{\Bbb R}^{M+1}$ (with
$\sum_{{t}_{i}}$
and $\sum_{{t}_{f}}$ as boundary), parametrized by $t$ and $(\varphi^r)_{r=1,
\cdot M}$, and with the euclidean metric
\begin{equation}
(G_{ij}^E)_{ij=1, \cdots, M+1}=\left(\begin{array}{cc}
\displaystyle g_{rs}=(\rho\cdot\cos\theta)^{\frac{2}{M}}{\bar g}_{rs} &
\displaystyle 0 \\
& \vdots \\
0 \cdots 0 &{\dot{\vec x}}^2=\sin^2\theta \end{array}\right) \qquad .
\end{equation}
Again one may choose $\rho(\varphi)=\lambda^M=\mbox{const}$, for
simplicity. As (34) contains the entire information about ${\Bbb M}$, the
minimal hypersurface equations should be equivalent to the flatness of ${\Bbb
M}_E$, i.e. the vanishing of the curvature-tensor
\begin{eqnarray}
R_{ijkl}^E & = &
\frac{1}{2}(\partial_{il}^2G_{jk}^E+\partial_{jk}^2G_{il}^E-
\partial_{ik}^2G_{jl}^E-\partial_{jl}^EG_{ik}^E)\\
\nonumber & + & (G^E)^{mn}\,(\Gamma_{m, il}\,\Gamma_{n, jk}-\Gamma_{m,
ik}\,\Gamma_{n, jl} )\qquad .
\end{eqnarray}
Due to the special form of the metric, cp. (34), one has (with
$M+1=:N;a,b,c=1 \cdots M$)
\begin{eqnarray}
\Gamma_{N, Na} & = & \sin\theta\cos\theta\partial_a\theta=-\Gamma_{a, NN}\\
\nonumber \Gamma_{N, ab} & = & -\frac{1}{2}{\dot g}_{ab}=-\Gamma_{a, Nb}\\
\nonumber \Gamma_{N, NN} & = & \sin\theta\cos\theta \; \;
{\dot \theta} \, , \,
\Gamma_{a, bc}=\gamma_{a, bc} \qquad \mbox{(cp. (14))} \qquad .
\end{eqnarray}
Using (36) one indeed finds the following: $R_{Nr, Ns}^E = 0 \qquad
\mbox{is equivalent to (19);}$
\begin{equation}
R_{Nabc}^E =  \frac{\sin\theta}{2}(\nabla_c(\frac{{\dot
g}_{ab}}{\sin\theta}-\nabla_b(\frac{{\dot g}_{ac}}{\sin\theta}))
\end{equation}
so that, by defining $h_{ab}$ according to (11), the vanishing of (37) is
equivalent ot the Codazzi-equations (27). Finally,
\begin{equation}
R_{abcd}^E=R_{abcd}-\frac{1}{4\sin^2\theta}({\dot g}_{ac}{\dot g}_{bd}-{\dot
g}_{ad}{\dot g}_{bc})
\end{equation}
so that the vanishing of (38) is equivalent to the Gauss-equations, (25). One
major advantage of this formulation is that the minimal hyper-surface
equations (due to the definition of the curvature tensor,
($\nabla_i\nabla_l-\nabla_l\nabla_i)x^j\equiv-R^E_{ilk}{^{j}}x{^k})$
are therefore the compatibility conditions
$([\partial_i+\Gamma_i,\partial_l+\Gamma_l]x^j=0)$ of the linear system of
equations
\begin{equation}
(\partial_i+\Gamma_i)\psi=0 \qquad i=1...M+1\qquad ,
\end{equation}
with $N\times N$ Matrices $(\Gamma_i)^j_k:=\Gamma^j_{ik}$.
Explicitely, one finds
\begin{equation}
\Gamma_c=\left(\begin{array}{ccc}
\displaystyle{\gamma^a_{cb}}
& \displaystyle{{\frac{1}{2}T_g^a{_c}}}\\
\\
\displaystyle{-\frac{{\dot g}_{ac}}{\sin^2\theta}}
&\displaystyle{\mbox{cot}\theta\partial_c\theta}
\end{array}\right)_{a,b,c,=1\cdots M}
\end{equation}
\medskip
\begin{equation}
\Gamma_N=\left(\begin{array}{ccc}
\displaystyle{\frac{1}{2}T_g}
&\displaystyle{-\sin\theta\cos\theta\partial^b\theta}\\
\\
\displaystyle{\mbox{cot}\theta\partial_b\theta}&\displaystyle{{\dot
\theta}\mbox{cot}\theta} \end{array}\right)
\end{equation}
\medskip
where $T_g$ and $\gamma^a_{cb}$ are as before (with
$g_{rs}=(\rho\cdot\cos\theta)^{\frac{2}{M}}{\bar g}_{rs}$), i.e. depending on
$\theta$ (and $\rho$) in the following way:
\begin{equation}
T_g={\bar T}-\frac{2}{M}\,{\dot \theta}\,\tan\theta\,\bf{1}
\end{equation}
\begin{equation}
\gamma^a_{bc}={\bar
\gamma}^a_{bc}+\frac{1}{M}(\delta_c^a\partial_b\ln(\rho\cdot\cos\theta)\, +
\,
\delta_b^a\partial_c\ln(\rho\cdot\cos\theta)-{\bar g}_{bc}{\bar
g}^{ad}\partial_d(\rho\cdot\cos\theta)) \qquad ,
\end{equation}
${\bar \gamma}^a_{bc}$ being the Christoffel-symbols corresponding to the
reduced metric ${\bar g}_{rs} ({\bar g}=1)$. For M=2, e.g., $\bar{g}_{ab}$
could be conveniently parametrized as
\begin{equation}
\bar{g}_{ab}=\left (\begin{array}{cc}
\displaystyle{\cosh\chi+\cos\phi\sinh\chi}
&\displaystyle{\sin\phi\sinh\chi}\\
\\
\displaystyle{\sin\phi\sinh\chi} &
\displaystyle{\cosh\chi-\cos\phi\sinh\chi}
\end{array}\right)\qquad .
\end{equation}

Considering
\begin{equation}
\phi^{(r)}(\varphi^1 \cdots\varphi^M, t) = \psi(\varphi^1, \cdots,
\varphi^{r}+\omega^r, \cdots\varphi^M,
t)\psi^{-1}(\varphi^1,\cdots, \; \varphi^r,\cdots,\; \varphi^M, t)
\;\; r=1 \cdots M
\end{equation}
($\psi$ the matrix of fundamental solutions of (39) and, for definiteness,
taking $\sum_M$ to be
an M-torus, with $\varphi^r\epsilon[0, \omega^r]$), satisfying
\begin{equation}
\partial_i\phi^{(r)}=[\phi^{(r)}, \Gamma_i]\qquad ,
\end{equation}
non-local conserved charges
\begin{equation}
Q_{rm}\,=\,Tr(\phi^{(r)})^m
\end{equation}
can be deduced from (39) -- expressable in terms of the Christoffel-symbols
$\Gamma^i_{jk}$ of ${\Bbb M}_E$ via solving (46)$_{i=r}$ as a pathordered
exponential,
\begin{equation}
\phi^{(r)}(\varphi^1\cdots\varphi^M, t)=\wp e^{-\int_{\varphi
r}^{\varphi{^{r}}+\omega{^{r}}}\Gamma_r(\varphi^1\cdots{\tilde
\varphi^r}\cdots\varphi^M, t)d{\tilde \varphi}}{^{r}}\qquad .
\end{equation}
It is extremely tempting to speculate that the hidden Lorentz-invariance
together with the $(S)\mbox{Diff}\sum$ invariance should allow one to
introduce a spectral parameter into (39). This would imply an infinity of
conserved quantities by expanding (47) in terms of this parameter
(note that the scale-parameter $\lambda$, cp (8), on which the $\Gamma_i$
at first sight seem to depend non-trivially, eventually just leads to a
conjugation of $\phi^{(r)}$ by a $\lambda$-dependent matrix).
\vskip .5truecm
\noindent{\bf{IV. Singularity Structure and Conserved Quantities for M=1
(Strings)}}
\vskip .5truecm
Due to the
fact that for $M=1$ eq. (5) (with $\rho=\lambda=\mbox{const}$) is trivial,
having
\begin{equation}
\vec x (t,\varphi)=\lambda(\,\vec a\, (\varphi+\frac{t}{\lambda})+\,\vec
b\,(\varphi-\frac{t}{\lambda}))
\end{equation}
as its general solution (with the components of $\vec a$ and $\vec b$ being
$2\pi$ periodic functions, for closed strings) the possible motions of the
string can be obtained explicitely, by inserting (49) into (6) and (7)
(yielding ${\vec a}'^2={\vec b}'^2=\frac{1}{4}$), so that
\begin{equation}
{\vec x}'(t,\varphi)=\lambda\cos\,(f-g)\left(\begin{array}{c}
\displaystyle{-\sin\,(f+g)}\\
\\
\displaystyle{\cos\,(f+g)}
\end{array}\right)
\end{equation}
\[
{\dot{\vec x}}(t,\varphi)=-\sin\,(f-g)\left(\begin{array}{c}
\displaystyle{\cos\,(f+g)}\\
\\
\displaystyle{\sin\,(f+g)}
\end{array}\right)
\]
where $f=f(\varphi+\frac{t}{\lambda})$ and
$g=g(\varphi-\frac{t}{\lambda})$; from now on,
$\lambda$ will be put equal to 1, for simplicity. Apart from the
requirement that (44) should describe a closed curve, $f$ and $g$ are
arbitrary. The closedness-condition is important, as it forbids, e.g., to
choose $f$ and $g$ to be small on the entire interval $[0,\,2\pi]$; moreover,
as the range of $f+g$ has to be at least $2\pi$, it is easy to see
that even if
$|f-g|<\pi$ initially, there will always exist a finite time $t_s$ at which
$|f-g|=\pi$ (and $f'+g'\neq0$) for some point on the string (i.e. some
$\varphi_s$). At $(t_s,\varphi_s)$ the worldsheet can not be regular -- the
curvature $k$ of the string diverges as
\begin{equation}
k(t,\varphi)=\frac{f'+g'}{\cos(f-g)}
\end{equation}
-- hence one finds that any (!) closed string motion in ${\Bbb R}^2$ (that
was supposed to extremize the area functional in Minkowski-space) must be
(become) singular. Infinitely extended regular ``minimal'' hypersurfaces in
${\Bbb R}^{1, 2}$ of the topological type $S^1\times{\Bbb R}$ can not
exist. This fact is known (see e.g. [14]) but not really well
known. Considering the fact
that in the case of membranes moving in ${\Bbb R}^3$ (i.e. 3+1 dimensional
space-time) it has often been argued (and taken against such theories) that
regular motions will not exist due to an impossibility of balancing the
surface tension by rotation the lack of thought concerning singularities in
string theories (which are rather commonly believed to be stabilized by
rotation) is somewhat astonishing -- in particular as these singularities
appear to be one of their interesting (rather then disturbing)
features. Due to (50) it is clear that for smooth $f$ and $g$ such
singularities not only appear, but also go away smoothly (i.e., in the context
of the orthonormal gauge, can be uniquely extended beyond the
singularity). There also exist choices for $f$ and $g$,
for which the number of
singularities is constant in time, e.g.
\begin{equation}
\vec x=\frac{1}{2m}\left(\begin{array}{c}
\displaystyle{\cos(m(\varphi+t))}\\
\\
\displaystyle{\sin(m(\varphi+t))}
\end{array}\right)+\frac{1}{2n}\left(\begin{array}{c}
\displaystyle{\cos(n(\varphi-t))}\\
\\
\displaystyle{\sin(n(\varphi-t))}
\end{array}\right)
\end{equation}
(m, n being two different
integers with no common divisor $\ne\pm1$). (52) coresponds to
choosing $f=\frac{m}{2}(\varphi+t), g =\frac{n}{2}(\varphi-t)$ in
(50), and describes a closed curve of time-independent shape, rotating with
constant angular velocity $\omega=|\frac{2mn}{m-n}|$ around the origin,
having $|m-n|$ cusps -  the minima of
\begin{equation}
|\vec x|=r({\tilde
\varphi}:=\varphi+\frac{m+n}{m-n}t)=\frac{(m-n)^2}{4m^2n^2}+
\frac{1}{mn}\cos^2(\frac{m-n}{2}{\tilde\varphi})
\qquad .
\end{equation}
Note that ${\tilde \varphi}$ does not coincide with the geometrical angle
arctan $\frac{x_2}{x_1}$, and that the curves (52) have the length
$L=\int_0^{2\pi}|{\vec x}'|d\varphi=4$, independent of $m$ and $n$ (and
$t$, of course).

In order to see why regular shapes can not be balanced by rotation, one can
insert the Ansatz
\begin{equation}
{\vec x}(t, \varphi)=e^{{1\,0\choose0\,1}f(t)}{\vec m}(\varphi)
\end{equation}
into the $\mu=0$ part of (3), giving up the orthogonality-condition (7); with
\begin{equation}
\sqrt{G}G^{\alpha\beta}=\frac{-1}{\sqrt{{\vec m}'^2(1-{\dot f}^2{\vec
m}^2)+{\dot f}^2(({\vec m}\times{\vec m}')^2}}\left(\begin{array}{cc}
\displaystyle{-{\vec m}'^2}&\displaystyle{{\dot f}({\vec m}\times{\vec m}')}\\
\displaystyle{{\dot f}({\vec m}\times{\vec m}')}&\displaystyle{1-{\dot
f}^2{\vec
m}^2} \end{array}\right) \quad ,
\end{equation}
${\vec m}\times{\vec m}':=m_1m_2'-m_2m_1'$, one gets
\begin{equation}
\partial_t\left(\frac{|{\vec m}'|}{\sqrt{1-{\dot
f}^2r^2\cos^2\theta}}\right)=\partial_{\varphi}\left(\frac{{\dot
f}r\sin\theta}{\sqrt{1-{\dot f}^2r^2\cos^2\theta}}\right)
\end{equation}
where $r=|{\vec x}|=|{\vec m}|$ and $\theta=\not\prec({\vec m},{\vec m}')$ are
functions of $\varphi$ (which for a regular curve could be chosen to be the
arclength, setting $|{\vec m}'|=1$). Dividing by ${\dot f}$ one gets
\begin{equation}
{\ddot f}r^2\cos^2\theta|{\vec m}'|+{\dot
f}^2r^3(\sin\theta)'=(r\sin\theta)' \qquad .
\end{equation}
Excluding the case $r\cdot\sin\theta=\mbox{const}$ (which, too, can not
correspond to a regular curve, s.b.) one finds $f(t)=\omega\cdot t$ (as
expected, due to the assumption of time-independent shape) and by integrating
(57) (or directly from (56))
\begin{equation}
\omega^2r^2\cdot(1+c\sin^2\theta)=1
\end{equation}
This yields the ``desired'' conclusion, as for $c<0\,(c>0) \sin^2\theta$ would
have to be minimal (maximal) when $r$ assumes it minimum (maximum). A rather
special class
of solutions consists of string-motions with ${\dot{\vec
x}}(t=0,\varphi)\equiv 0$; from (50) it is clear that $f\equiv g=:\frac{h}{2}$
in this case. As an example, consider the following ``harmonic perturbations
of the radially symmetric string solution'':
\begin{equation}
h(\varphi)=\varphi+\,\epsilon\,\sin(m\varphi)_{m\,\epsilon\,{\Bbb Z}}
\end{equation}
\begin{equation}
{\vec x}'(t,\varphi)=\cos(t+\epsilon\sin mt \, \cos
m\varphi)\cdot{-\sin\,(\varphi+\epsilon\sin m\varphi\cos
mt)\choose\cos\,(\varphi+\epsilon\sin m\varphi\cos mt)}
\end{equation}
\bigskip
\begin{equation}
k(t,\varphi)=\frac{1+m\epsilon\cos m\varphi\cos mt}{|\cos\,(t+\epsilon\sin mt
\cos m\varphi)|}
\end{equation}
(so far, no approximation was made). Suppose now, that $|m\epsilon|<<1$; the
two cases I) $m$ odd, II) $m$ even show drastically different behaviour. Case
I: (60) becomes singular shortly before $t=\pi/2$, \, at \, $m$ (equally
distributed) discrete points $\varphi_i$; for a small time-interval these $m$
singularities move ``along
the string'' (in particular, disappear instantaneously at the $\varphi_i$);
shortly after $t=\pi/2$ all singularities disappear, and at $t=\pi$ the string
is back to its original shape (with $\varphi \to \varphi+\pi$) and at rest.
Case II: as $\sin mt\to 0$ for even $m$, $(t\to\pi/2)$,
the string stays regular
until $t=\pi/2$ (when it has shrunk to a point), grows again and reaches its
initial conditions at $t=\pi$.

It would be interesting to use the infinitely
many known conservation laws for the string, including e.g. non-local charges
[7]
\begin{equation}
Q_{\mu{_1}\cdots\mu{_n}}^{\pm}=\int_{\varphi}^{\varphi+2\pi}
d\varphi_1\int_{\varphi}^{\varphi{_1}}d\varphi_2 ...
\int_{\varphi}^{\varphi{_{n-1}}}d\varphi_{n}
\, u_{\mu{_{1}}}^{\pm}\cdot u_{\mu{_{2}}}^{\pm}\cdot ... \cdot
u_{\mu{_n}}^{\pm}\mbox{+ all cyclic permutations}
\end{equation}
$(u_0^{\pm}=1 \, , \, {\vec u}^{\pm}={\dot{\vec x}} \pm {\vec x}')$ and
non-polynomial ones (going back to [8]),
\begin{equation}
Q_{F,G}=\int\frac{dx}{2{\dot p}+p'^2}\cdot\{\frac{1}{F'(p'+\sqrt{2{\dot
p}+p'^2}}+\frac{1}{G'(p'-\sqrt{2{\dot p}+p'^2})}\}
\end{equation}

(where $p=x^0-x^2$, expressed as a function of $x=x^1$ and
$\tau=\frac{x^0+x^2}{2};
\cdot = \frac{\partial}{\partial\tau}, '=\frac{\partial}{\partial x}$, $F$ and
$G$ arbitrary functions) for an understanding of the string motion,
e.g. (should the curve be star-shaped around the center of mass) in the
radial (non-parametric) representation, $r=r(t,\varphi=$polar angle),
\begin{equation}
H=\int_0^{2\pi}\sqrt{1+\frac{p^2}{r^2}}\sqrt{r^2+r'^2}d\varphi\qquad ,
\end{equation}
with equations of motion
\begin{equation}(\cdot=\frac{\partial}{\partial
t},'=\frac{\partial}{\partial\varphi}) \, , \, {\dot
r}=\frac{p}{r}\sqrt{\frac{r^2+r'^2}{p^2+r^2}} \, , \, {\dot
p}=-\sqrt{\frac{r^2+r'^2}{p^2+r^2}}+\frac{1}{r}(r'\sqrt{\frac{p^2+r^2}
{r^2+r'^
2}})'
\end{equation}
respectively
\begin{equation}
{\ddot r}(r^2+r'^2)-r''(1-{\dot r}^2)-2{\dot r}r'{\dot r}'+r(1-{\dot
r}^2)+\frac{2r'^2}{r}=0
\end{equation}
(note that $\sqrt{r^2+r'^2}d\varphi$ is the infinitesimal arc-length of the
curve, and that in (66) the combination of terms not involving
time-derivatives of $r$ is proportional to the curvature). Simple properties
may be directly deduced from the time-independence of $H$, e.g.: As
$p/r=\frac{{\dot
r}}{\sqrt{1-{\dot r}^2+\frac{r'^2}{r^2}}}\to
0 \; \mbox{when}\frac{r^2}{r'^2}\to 0$ (implying
$\int_0^{2\pi}d\varphi|r'|\to\mbox{const}$, i.e. the string becoming
infinitely rough, if $r\to0$), but $\sqrt{1+\frac{r'^2}{r^2}}$
finite otherwise
(implying $p=\frac{r{\dot r}}{\sqrt{(1+\frac{r'^2}{r^2})-{\dot r}^2}}\to 0$
for $r\to 0, {\dot r}^2-1 \, not \to \, 0$) one finds that in the latter
case one
actually must have $\frac{r'}{r}\to0$ and ${\dot r}^2\to 1$, i.e. the
singularity being light-cone like.
\vskip .5truecm
\noindent{\bf{V. Some Explicit Hypersurface Solutions}}
\vskip .5truecm
In addition to the methods described in
[9], solutions (of (3)) of the following form may be found:
\begin{equation}I) \sum^{M+1}_{\mu=0}f_{\mu}(x^{\mu})=0
\end{equation}
\[ II)\; (\vec x -\vec a(x^0))^2-r^2(x^0)=\mbox{const}\qquad .
\]
In both cases it is easiest to use that the level sets $u=\mbox{const}$
(cp. [3]) of functions $u(x^0 \cdots x^{M+1})$ satisfying
\begin{equation}
(\eta^{\mu\nu}\eta^{\rho\lambda}-\eta^{\mu\rho}\eta^{\nu\lambda})
\frac{\partial u}{\partial x^{\mu}}\frac{\partial u}
{\partial x^{\nu}}\frac{\partial^2u}{\partial x^{\rho}\partial
x^{\lambda}}=0
\end{equation}
are extremal hypersurfaces in ${\Bbb R}^{1, M+1}$.

\noindent Ansatz I:
\noindent Considering $F_{\mu}:=(f'_{\mu})^2$ as functions of
$v_{\mu}:=f_{\mu}(x_{\mu})$ (implying
$f''_{\mu}=\frac{1}{2}F'_{\mu}(v_{\mu})$, where ' always means derivative with
respect to the relevant variable)
\begin{equation}
{f'_0}^2 {\sum^M_{i=1}}{f''_i}+{f_0}''{\sum^M_{i=1}}{f_i}'^2=
{\sum_{i\ne j}}{f_o}'^2{f_j}''
\end{equation}
becomes a first order (functional-) differential equation,
\begin{equation}
F_0\sum^M_1 F_i'+F_0'\sum_1^M F_i=\sum_{i\ne j}F_iF_j'
\end{equation}
which is to be solved on the constraint surface $\sum^{M+1}_{\mu=0}\,
v_{\mu}=0
(\mbox{so}\, F_0' (v_0)=-F_0'\,^{\mu=0}(\sum^M_1 v_i)$ if $F_0$ is an even
function of $v_0$).
Depending on the dimension, $M$, (70) may admit ``soliton'' solutions of
different type. While for $M=1$ there are solutions of the form $(a_{\mu}\ne
0)$
\begin{equation}
F_{\mu}=a_{\mu}+b_{\mu}e^{\kappa v{_{\mu}}}+c_{\mu}e^{-\kappa v{_{\mu}}}
\end{equation}
the constant term has to be zero for $M=2$, where particular solutions are
\begin{equation}
F_{\mu}=\frac{-\epsilon^2}{c{_{\mu}}}e^{\kappa v{_{\mu}}}+c_{\mu}e^{-\kappa
v{_{\mu}}}>0
\end{equation}
\[c_0 \cdot c_1 \cdot c_2 \cdot c_3 = -\epsilon^4\qquad .
\]
Defining $h$ as $\frac{c_{\mu}}{4}e^{-\kappa v{_{\mu}}}$ if $c_{\mu}>0$,
respectively $\frac{-\epsilon^2}{4c_{\mu}}e^{\kappa v{_{\mu}}}$ for
$c_{\mu}<0$, yielding $\frac{h'^2}{\kappa^2}=4h^3-\frac{\epsilon^2}{4}h$,
i.e. (irrespective of the choice of $c_{\mu}$) the elliptic
Weierstrass-function $h{(w)}=\wp(\kappa w+w_0;g_2=\frac{\epsilon^2}{4},
g_3=0)=\frac{\epsilon}{4}\wp(\sqrt{\frac{\epsilon}{4}}(\kappa w+w_0);g_2=4,
g_3=0)$, cp [10], -- one finds (using $\sum f_{\mu}=0$) that up to
scale
transformations $x^{\mu}\to\lambda x^{\mu}$, translations
$x^{\mu}\to x^{\mu}+d^{\mu}$, Lorentz transformations
$x^{\mu}\to\wedge^{\mu}_\nu x^{\nu}$, and permutations of the spatial
coordinates one gets two inequivalent solutions, namely
\begin{equation}
\wp(x)\,\wp(y)\,\wp(z)\,=\,\wp(t)
\end{equation}
and
\begin{equation}
\wp(x)\,\wp(y)\,\wp(t)\,=\,\wp(z)
\end{equation}
with the invariants $g_2$ of the $\wp$-functions equal to 1 (hence all
$\wp$'s having period
$2\omega=\sqrt{2}K(\frac{1}{\sqrt{2}})=\frac{1}{2\cdot\sqrt{2\pi}}
(\Gamma(\frac{1}{4}))^2$
and taking the value 1 as their minimum). Before discussing the solutions (73)
and (74) as time-dependent surfaces $\sum_t$ in ${\Bbb R}^3$, it seems
worthwhile to note how they evolve from various other points of view. E.g.,
replacing the Ansatz I) by the (equivalent) Ansatz
\[
{\tilde I})\, u(x^0,\cdots,x^{M+1})={\Bbb T}_{\mu=0}^N
\;\;g_{\mu}\;(x^{\mu})(=1)
\,\,
,\]
to be inserted into (68), i.e.$\sum\limits_{\mu\ne\rho}u^{\mu}
u_{\mu}u^{\rho}_{\rho}-2\sum\limits_{\mu<\nu}u^{\mu}u^{\nu}u_{\mu\nu}=0$,
one is led to a first order (functional) differential equation by taking the
unknown functions $g_{\mu}(x^{\mu})=:z_{\mu}$ as independent variables, and
the square of their derivatives, $g_{\mu}'^2=G_{\mu}(z_{\mu})$ as unknown
functions of the new variables. While the form of the equation to be solved on
the constraint surface $z_0z_1\cdots z_N=1$ is slightly more involved than
(70), one may now look for polynomial solutions. (73), e.g., respectively
\begin{eqnarray}
f_0(x^0) & = & -\ln\wp(x^0)\\
\nonumber f_i(x^i) & = & +\ln\wp(x^i)=:X{_{i}}\;\;{_{i=1, 2, 4}}
\end{eqnarray}
corresponds to
\begin{equation}
G_i(z)=-G_0(z)=4z(z^2-1)=:G(z)
\end{equation}
satisfying
\begin{equation}
\frac{1}{2}\,(z_0 G\,(z_0)z_1^2G'\,(z_1)z_2^3z_3^3+11 \,\mbox{more
terms})-2\,(z_0G\,(z_0)z_1G\,(z_1)z_2^3z_3^3+5 \,\mbox{more})\,\equiv\, 0
\end{equation}
on $z_0z_1z_2z_3\,\equiv\,1$. In this ``derivation'', the fact that (up to a
sign), the Weierstrass $\wp$-function satisfies the same differential equation
as its inverse, $\frac{1}{\wp}$, plays a crucial role (otherwise, the
resulting equation would be much more complicated, than (77)).

It is easy to
deduce from (73), (74) that in the first case, all points of the surface
$\sum_t$ always move with a velocity $\ge$ 1 (=1 if and only if at least 2 of
the spatial coordinates are equal to $\omega$ mod $2\omega$), whereas with a
velocity $\le 1$ (=1 if and only if $x=\omega=y$ mod $2\omega$) in the latter
case. Hence (73) defines a space-like maximal hypersurface of ${\Bbb R}^{1,
3}$; it provides e.g. a nice example in the context of theorems on isolated
singularities (of area-maximizing hypersurfaces)[11] and generalized
Bernstein-theorems [12]. (73) may be described as follows (restricting to one
unit cube $C=\{ {\vec x}\epsilon{\Bbb R}^{3}|x,y,z\;\epsilon[0, 2\omega]\}$:
At $t=\omega$, (73) implies $x=y=z=\omega$ (a point); at $t=0=2\omega$,
$\sum_t$ consists of all faces of $C$. When $t$ varies from 0 to $\omega$, the
-- initially square-surface $\sum_t$ becomes rounder (always convex), finally
vanishing as a round point, as can easily be seen by expanding $\wp$ around
$\omega$, yielding in first order the light cone
\begin{equation}
(x-\omega)^2+(y-\omega)^2+(z-\omega)^2=(t-\omega)^2
\end{equation}
as $t \to\omega$. As $t$ varies from $\omega$ to $2\omega$, the reversed
picture holds (the surface becoming more square, due to the fact that the
further $x$ and $y$, e.g., are away from $\omega$, the faster $z(x,y)$ moves
up, respectively down).

The mean-respectively Gauss-curvature of $\sum_t$ are
\begin{equation}
H=\frac{1}{(X'^2+Y'^2+Z'^2)^{3/2}}\cdot\left(
X''(Y'^2+Z'^2)+Y''(X'^2+Z'^2)+Z''(X'^2+Y'^2)\right)
\end{equation}
\begin{equation}
K=\frac{R}{2}=\frac{1}{(X'^2+Y'^2+Z'^2)^{2}}\cdot\left(
X''Y''Z'^2+X''Z''Y'^2+Y''Z''X'^2\right)
\end{equation}
where $X=\ln\wp(x),\cdots$, (cp. 75), and $X+Y+Z=\ln\wp(t)=\mbox{const}$ (on
each $\sum_t$). From $\wp$ having a second order pole at 0 (with residue 1)
one can easily find the exact form of the curvatures at those points of
$\sum_t$ that approach the corners, resp. edges, of $C$.

Solution (74), on the other hand (again restricting to one unit cube $C$), is
such that the upper and lower edges of $C$ always belong to $\sum_t$, acting
like a fixed frame. At $t=0$ $\sum_t$ is flat (covering the upper and lower
face of $C$); as t grows, so do the upper part of $\sum_t$ (moving downwards)
as
well as the lower part of $\sum_t$ (moving upwards); at $t=\omega$, the two
parts touch at $x=y=z=\omega$, where the curvatures diverge. For $t>\omega$,
the
process reverses $(\sum_{\omega+{\tilde t}}=\sum_{\omega-{\tilde t}})$. In
order to find solutions of the form (67)$_{II}$, one inserts $u({\vec
x},t)=\frac{1}{2}({\vec x}-{\vec a}(t))^2-\frac{1}{2}r^2(t)$ into (68). With
${\vec \nabla u}=({\vec x}-{\vec a}), ({\vec \nabla}u)^2=r^2(t)+2u, {\vec
\nabla}^2u=N, {\vec \nabla}{\dot u}=-{\dot{\vec a}}$ one can write the
resulting equation in the form
\begin{equation}
{\dot F}+(N-2)F^2=\frac{N-1}{r^2(t)}\qquad ,
\end{equation}
where
\begin{equation}
F\,:\,=\,\frac{{\dot u}}{r^2(t)}\qquad .
\end{equation}
When $N=2$, (81) implies
\begin{equation}
F(t,{\vec x})=f({\vec x})+\int_0^t\,\frac{d{\tilde t}}{r^2({\tilde t})}
\end{equation}
while a comparison of ${\dot u}={\dot{\vec a}}({\vec a}-{\vec x})-r{\dot r}$
with (82) then yields
\begin{equation}
f({\vec x})={\vec \lambda}\cdot{\vec x}\, , \, {\vec a}(t)={\vec a}_0-{\vec
\lambda}\int_0^tr^2({\tilde t})d{\tilde t}
\end{equation}
\begin{equation}
{\vec \lambda}\int_0^tr^2-{\vec \lambda}\cdot{\vec a}_0-({\dot
{\ln}}r)-\int_0^t\frac{1}{r^2}=0\qquad .
\end{equation}
Differentiating (85), putting $\ln r^2=h(t)$, multiplying by ${\dot h}$,
integrating and letting $h=-\ln H$, one finds that up to an additive constant
$\mu, H$ equals the Weierstrass $\wp$-function (with $g_2=12\mu^2-4{\vec
\lambda}^2$ and $g_3=-4\mu(\mu^2+{\vec \lambda}^2)$).

So
\begin{equation}
r(t)=\frac{1}{\sqrt{\wp(t)+\mu}} \qquad .
\end{equation}

This solution was also found in [13].

\vskip 1truecm
\noindent{\bf{Acknowledgement}}
\vskip .5truecm
I would like to thank A. Bobenko, M. Bordemann, A. Chamseddine, J. Fr\"ohlich,
Y. Giga,
K. Happle, G. Huisken, M. Struwe, S. Theisen and K. Voss for helpful
discussions, E. Heeb for a drawing of one of the curves (52),
N. Bollow for visualizing solution (73) on a computer, the Deutsche
Forschungsgemeinschaft for financial support, and the Institute for
Theoretical Physics of the Swiss Federal Institute of Technology, Z\"urich, as
well as the mathematics department of T\"ubingen- and Hokkaido University, for
hospitality.
\vfill\eject
\noindent{\bf{References}}
\begin{itemize}
\item[[1]] J. Nambu; Copenhagen Summer Symposium 1970, unpublished.
\item[] T. Goto; Progr. Theor. Physics 46 (1971) 1560.
\item[[2]] J. Hoppe; MIT Ph.D. Thesis (1982) [Elem. Part. Res. J. (Kyoto) 80
(1989)145].
\item[[3]] M. Bordemann, J. Hoppe; Phys. Lett B 325 (1994) 359.
\item[[4]] J. Hoppe; ``Canonical 3+1-Description of Relativistic Membranes'',
Yukawa Institute preprint, K-1097-94.
\item[[5]] J. Hoppe; Phys. Lett. B 335 (1994) 41.
\item[[6]] G. Huisken; J. of Diff. Geometry 20 (1984) 237.
\item[[7]] K. Pohlmeyer; Phys. Lett. B 119 (1982) 100.
\item[[8]] J.M. Verosky; J. Math. Phys. 27 (1986) 3061.
\item[[9]] J. Hoppe; Phys. Lett. B 329 (1994) 10.
\item[[10]] I.S. Gradsteyn, I.M. Ryzhik; ``Tables of Integrals, Series and
Products'', Academic Press 1965.
\item[[11]] K. Ecker; Manuscripta Math. 56 (1986) 375.
\item[[12]] S.Y. Cheng, S.T. Yau; Annals of Mathematics 104 (1976) 407.
\item[[13]] K. Happle, T. Kornhass; ``Klassische Zyklische Strings im
3-dimensionalen Minkowski-Raum'', Freiburg preprint, Oct. 1990, unpublished.
\item[[14]] G.P. Pron'ko, A.V. Razumov, L.D. Solov'ev; Sov. J. Part. Nucl. 14
(3)(1983) 229.
\end{itemize}
\end{document}